\documentclass[lettersize,journal]{IEEEtran}
\IEEEoverridecommandlockouts
\usepackage{array}
\usepackage{amssymb,amsmath,graphicx,hyperref,lineno,verbatim}
\usepackage{xcolor}
\usepackage[noadjust]{cite}
\usepackage{multirow}
\usepackage{fancyhdr}
\title{UHF RFID and NFC Point-of-Care -- Architecture, Security, and Implementation}

\author{Giulio Maria Bianco, Emanuele Raso, Luca Fiore, Vincenzo Mazzaracchio, Lorenzo Bracciale, Fabiana Arduini, Pierpaolo Loreti, Gaetano Marrocco, and Cecilia Occhiuzzi}

\date{%
    $^1$Organization 1\\%
    $^2$Organization 2\\[2ex]%
}

\begin{document}

\maketitle
\begin{abstract}
Points-of-care (PoCs) augment healthcare systems by performing care whenever needed and are becoming increasingly crucial for the well-being of the worldwide population. Personalized medicine, chronic illness management, and cost reduction can be achieved thanks to the widespread adoption of PoCs. Significant incentives for PoCs deployment are nowadays given by wearable devices and, in particular, by RFID (RadioFrequency IDentification) and NFC (Near Field Communications), which are rising among the technological cornerstones of the healthcare internet of things (H-IoT). To fully exploit recent technological advancements, this paper proposes a system architecture for RFID- and NFC-based PoCs. The architecture comprises in a unitary framework both interfaces to benefit from their complementary features, and gathered data are shared with medical experts through secure and user-friendly interfaces that implement the Fast Health Interoperability Resource (FHIR) emerging healthcare standard. The selection of the optimal UHF and NFC components is discussed concerning the employable sensing techniques. The secure transmission of sensitive medical data is addressed by developing a user-friendly "PoC App" that is the first web app exploiting attribute-based encryption (ABE). An application example of the system for monitoring the pH and cortisol levels in sweat is implemented and preliminarily tested by a healthy volunteer.
\end{abstract}
\section{Introduction}\label{sec:introduction} 
\begin{figure}
    \centering
    \includegraphics[width=\linewidth]{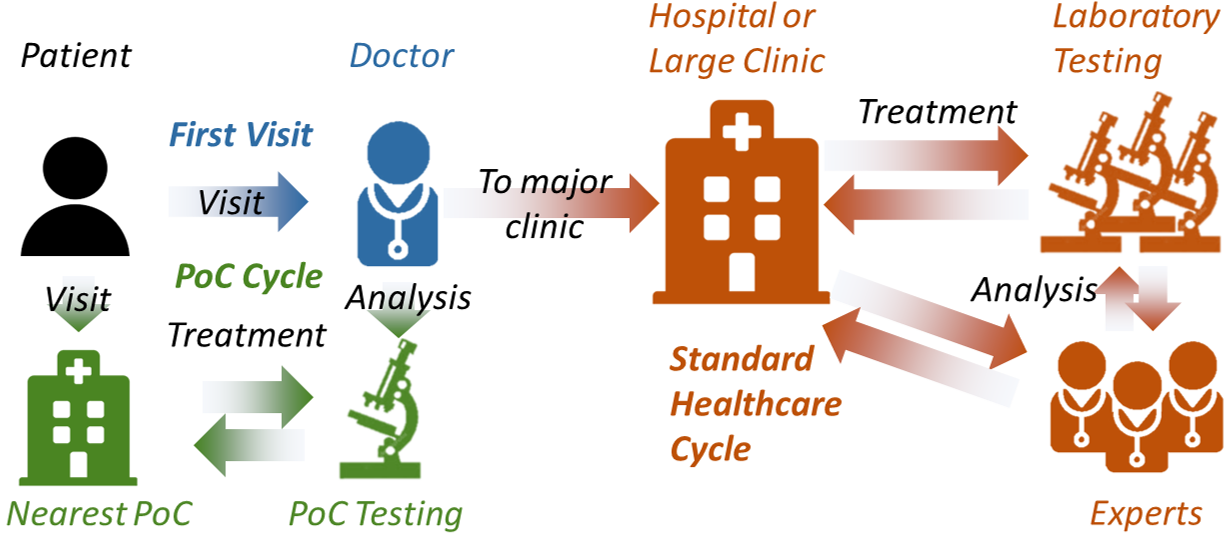}
    \caption{Concept of a \textit{PoC cycle} augmenting an existing healthcare system and timely treating the patient outside principal laboratories.}
    \label{fig:PoCsHealthcare}
\end{figure}

\let\thefootnote\relax\footnotetext{Giulio Maria Bianco, Gaetano Marrocco, and Cecilia Occhiuzzi are with the Pervasive Electromagnetics Laboratory of the University of Rome ``Tor Vergata"\par Emanuele Raso, Lorenzo Bracciale, and Pierpaolo Loreti, are with the University of Rome ``Tor Vergata" \par Luca Fiore, Vincenzo Mazzaracchio, and Fabiana Arduini are with the NanoBioSensing Laboratory of the University of Rome ``Tor Vergata".\par Work funded by Regione Lazio, project E-CROME (biosEnsori su Carta wiReless per la telemedicina in Oncologia e la misura di eMocromo ed Elettroliti; Development of NFC interface sensors for the measurement of biomarkers in blood), CUP: E85F21001040002.\par Contact author: giulio.maria.bianco@uniroma2.it}
%
%
Since $1990$, steady-paced medical advancements have enlengthened life expectancy worldwide; however, ageing, disability, and chronic illnesses have yielded a heavier disability burden on the population~\cite{Kassebaum2022} and, consequently, more expenses for healthcare systems~\cite{Bloom2020}. To manage the increasingly common chronic medical conditions, cost-effective and continuous treatments are needed. In this context, points-of-care (PoCs) allow for decreasing costs while raising the quality of medical care. PoCs are defined as sites of patient care wherever the care is performed and encompass both testing and monitoring~\cite{Kost21}. Since PoCs operate outside the main laboratories~\cite{Kost08}, they augment healthcare infrastructures by providing more frequent feedback loops closer to the patients, thus enabling precise, timely diagnoses and personalized treatments while lowering the costs~\cite{Campuzano21,Rasooly06} (Fig.~\ref{fig:PoCsHealthcare}). In addition to more conventional devices, the myriad of up-to-date data gathered through wireless sensors~\cite{Salimiyanrizi22, Merazzo21} can even support the integration of PoCs with the latest paradigm in medicine of \textit{homespitals}~\cite{Bloem20} and \textit{expert patients}~\cite{AnampaGuzman22}.\par
\begin{table}[t]
    \centering
    \caption{Examples of UHF and NFC sensors useful for PoC.}
    \label{tab:bibliographicTable}
    \begin{tabular}{l|l|l|l}
         &\textbf{UHF}&&\\
         \textbf{Ref.} & \textbf{or NFC} & \textbf{Sensing} & \textbf{Description}\\
         \hline\hline
         \cite{Camera20}& UHF & Physical & Device for fever monitoring\\
         \hline
         \cite{Jiang20}& NFC & Physical & Textile sensor for skin's temperature\\
         \hline
         \cite{Amendola15}& UHF & Behavioural & Movements discrimination \\
         \hline
         \cite{Qin11}& UHF & Behavioural  & Supervising care administration\\
         \hline
         \cite{Rafferty17}& NFC & Behavioural &  Annotating events' location via app\\
         \hline
         \cite{Mazzaracchio21}& UHF & Chemical &Checking the pH of sweat\\
         \hline
         \cite{Barba22}& NFC & Chemical &Monitoring cortisol in sweat\\
         \hline
         \cite{Promsuwan23}& NFC &  Chemical &Potentiostat for glucose sensing\\
    \end{tabular}
\end{table}
\begin{figure*}
    \centering
    \includegraphics[width=18cm]{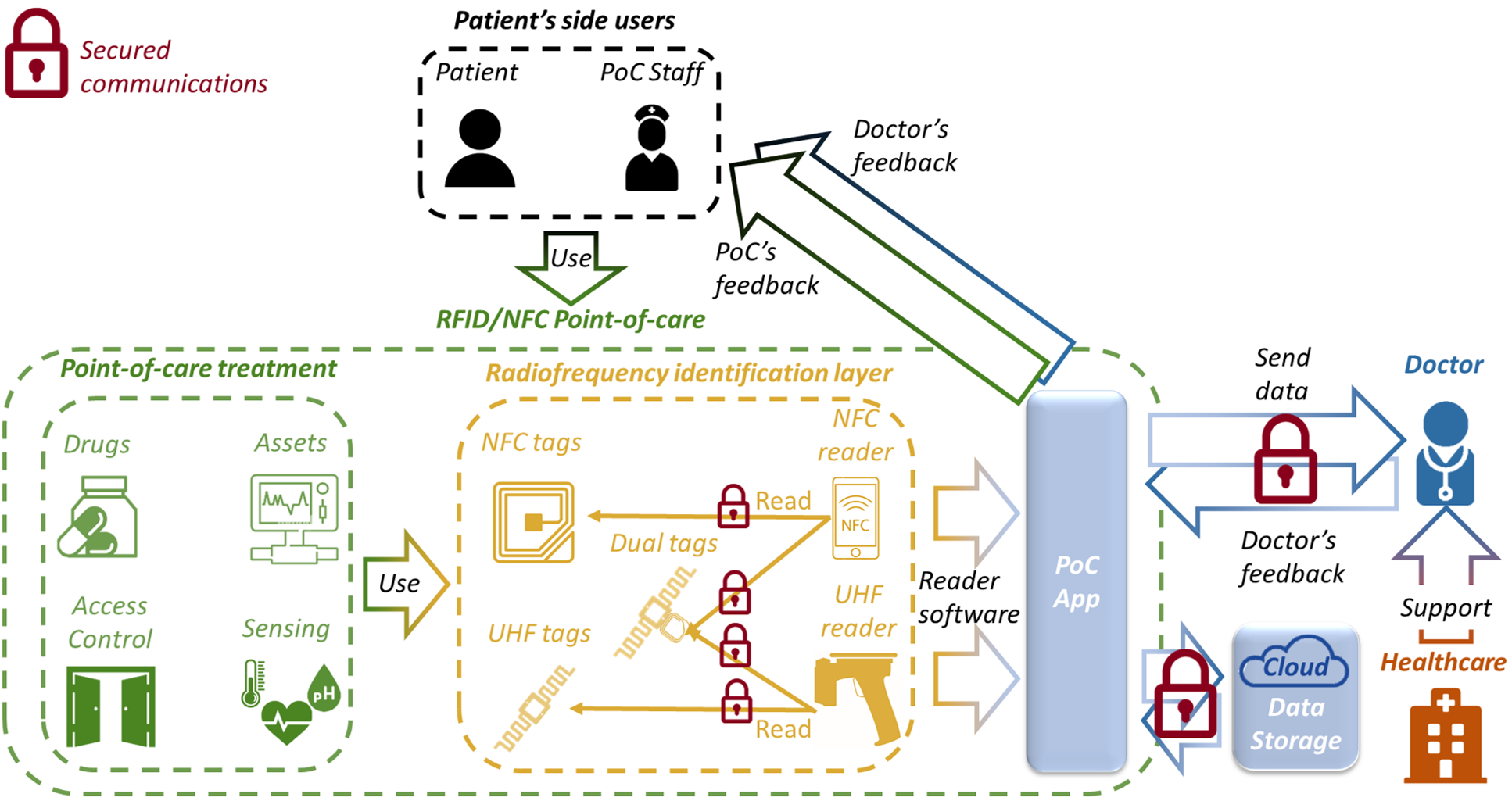}
    \caption{Proposed architecture of the UHF-RFID/NFC-PoC platform and its logical blocks.}
    \label{fig:architecture}
\end{figure*}
Among the available technologies, PoCs can also exploit ultra-high frequency (UHF) RFID (RadioFrequency IDentification) and NFC (Near Field Communication). Indeed, in the last decade, such devices have quickly arisen as versatile enablers of healthcare internet of things (H-IoT) systems by making processes more efficient~\cite{Yao12} and implementing pervasive monitoring of health status~\cite{Bianco21ASurvey}.
A straightforward use of RFID and NFC is augmenting PoCs with wireless identification to improve asset management~\cite{Smith19}, but more advanced tags can perform even complex sensing. In this latter case, the sensors included in the tags are low-power and low-cost to enable near-patient monitoring and testing. Last-generation sensor tags can sense many on- and off-body phenomena, including variations of temperature~\cite{Jiang20}, humidity~\cite{Eldebiky18}, pressure~\cite{Miozzi21Zero} and light~\cite{Wagih22}. Table~\ref{tab:bibliographicTable} reports examples of the most assessed sensing helpful for PoCs that can be categorized into three types: \textit{i})~physical, \textit{ii})~behavioural, and \textit{iii})~chemical.\par
As an example of physical sensors, temperature sensing allows for monitoring, directly and indirectly, a number of biological conditions and psycho-physical parameters, such as skin temperature~\cite{Panunzio21Second}, breath rate~\cite{Miozzi21Dual}, and inflammatory events~\cite{Avaltroni21}. Information for recognizing the behaviour of the patient or the medical staff can be obtained by knowing the reader's location and simply controlling when and where the tags are read~\cite{Qin11,Ohashi10} or, in a finer way, by feeding machine learning algorithms with the RSSI (received signal strength indicator) and phase recorded from the tags' responses~\cite{Amendola15}. Finally, chemical sensing is the most variegated type of sensing. Researchers are focusing on miniaturized chemical (bio)sensors for fast and on-site analysis of several clinically relevant biomarkers in body fluids like blood, sweat, and saliva by quantitative and accurate electrochemical techniques, i.e., potentiometry and amperometry~\cite{Emaminejad17, Gao16}. Only in the last decade such miniaturized chemical biosensors have been combined with transducers assisted by wireless technology, including UHF RFID~\cite{Mazzaracchio21,Xiao15,Rose15} and NFC~\cite{Barba22,Mirzajani22,Bandodkar19} since these technologies provide low operating costs, swift data transferring, and compatibility with flexible substrates~\cite{Kassal18}.\par
%
%
%
The listed examples suggest that radiofrequency identification devices could get medical data on the health status of a patient; however, to implement a real RFID/NFC-based PoC, many architectural and operative elements are still missing. Particularly, NFC and UHF RFID have different strengths and weaknesses, and since they require different interrogating devices, they are usually not combined. UHF RFID is used if a long reading distance, simultaneous reading of multiple tags, and/or ad-hoc antenna design due to size constraints are required. Using UHF hardware is also optimal if already available reading infrastructures, typically implemented for inventory purposes, can be exploited. On the contrary, NFC devices are deployed if near-contact distances, higher power and/or the highest security and data rates are required. Possible interoperability issues (like missed readings or hindered communications) between the NFC protocols are another weakness of the technology~\cite{Erb21}. In PoC scenarios and especially in domestic environments, the simultaneous use of UHF RFID and NFC devices can allow, thanks to their partial complementarity, for maximizing the benefits of the superior read range of the former and the close-range interaction of the latter in order to meet more patient's needs through the same platform. Indeed, a hybrid platform can provide better information redundancy in case of failure of one of the two reading architectures and can even be scaled up more effectively since they can be integrated with existing radiofrequency identification systems.\par
Besides the generation of data through sensors, PoCs also require an efficient transmission of data to the physician. Secure information sharing between patients and healthcare facilities/staff has not received sufficient attention from security experts yet, and patients often still send their sensitive data using e-mail or instant messaging services~\cite{manji2021using}. RFID and NFC sensors can generate huge amounts of data worsening the challenge. General-purpose Cloud services have then to be adopted so that two issues related to the privacy of the patient arise: \textit{i})~compliance of the monitoring platform with the regulatory framework on sensitive data, e.g., the European General Data Protection Regulation~\cite{Plug22}; \textit{ii})~adoption of proper technical solutions that ensure secure access to the data with no participation of the Cloud service, that, in general, could be an \textit{honest-but-curious}, or even \textit{malicious}, actor. For this reason, many solutions have been proposed in literature trying to deal with the major issue of ensuring privacy and data security in Cloud services~\cite{tang2016ensuring,samanthula2012efficient,maurer1996modelling}. A promising technology in this area is attribute-based encryption~(ABE), which offers cryptographic protection of information in combination with data access control directly provided by the technology and managed by data owners~\cite{li2010securing, hamsanandhini2022health, fabian2015collaborative, akinyele2011securing}.\par
Building on the previous considerations, this paper expands the preliminary work presented in~\cite{Bianco22Towards} and proposes a complete system architecture for PoCs based on radiofrequency identification for the first time. The architecture includes all the components, from the sensors to the data sharing, and seamlessly integrates UHF RFID and NFC in the same platform thanks to a novel web app. The web app encrypts data through a JavaScript library that wraps an ABE library written in Rust~\cite{RUSTRABE} and then transmits files according to the Fast Health Interoperability Resources (FHIR) emerging standard for healthcare files~\cite{saripalle2019using}. An implementation example simulating a PoC application exploits recent epidermal boards (from \cite{Fiore23,Barba22,Nappi21,Mazzaracchio21}) and shows how to deploy the system architecture.\par
The paper is organized as follows. The PoC system architecture and how to integrate all the necessary components are discussed in Section~\ref{sec:architecture}. We describe how the security issues can be addressed and implement an innovative web app which, to the best of our knowledge, is the first web app exploiting ABE (Section~\ref{sec:app}). Finally, Section~\ref{sec:implementation} shows an implementation of the architecture concerning the monitoring of sports activity by sensing temperature, pH, and cortisol.
\section{Architecture and Deployment}\label{sec:architecture}
\subsection{System Architecture Overview}\label{subsec:architecture}
Fig.~\ref{fig:architecture} depicts the system architecture of the UHF-RFID/NFC PoC. The \textit{patient's side users} are the patient himself/herself and any eventual medical staff assigned to the PoC; these users generate the medical data, transmit them to the doctor, and receive the doctor's feedback. The doctor following the patient's medical history, in turn, uses the PoC platform to receive the gathered data and, hence, provide the patient with personalized treatment or even change the care plan if needed.\par
%
To perform the required treatment, the RFID/NFC-PoC exploits a suitable combination of UHF RFID and NFC hardware. The hardware gathers data from medical care thanks to the tags and the readers that together compose the \textit{radiofrequency identification layer}. The healthcare information that can be processed by the RFID layer can be categorized into three main kinds~\cite{Bianco21ASurvey}: \textit{i})~drugs and assets management, \textit{ii})~access control, and~\textit{iii}) sensing, including environmental and behavioural sensing.\par
The operating system piloting the reader also runs an ad hoc \textit{PoC App}. Since the PoC exploits a hybrid UHF-NFC radiofrequency identification layer, the web app will run on both the typologies of readers to gather all the data. After the collection, the PoC App provides one first prompt \textit{PoC's feedback} to the patient's side users, informing them if the care is completed correctly and if any urgent action is needed to compensate for dangerous physiological alterations, for instance, adverse effects to drugs. The reader device formats the medical data to be seamlessly integrated with the information flux about the medical history of the patient in the computer healthcare system. Then, the PoC App encrypts data with a personal key of the patient and completes \textit{Data Storage} on a Cloud. Even commercial or personal Clouds like Google Drive or Dropbox can be exploited, given that files are already encrypted before upload. After data storage, the PoC App notifies the doctor that new data are available. Via the PoC App itself, the doctor can access the encrypted data on the Cloud, store a copy on a personal device, decrypt the copy using a personal key and, finally, access the medical information.\par
At this point, the doctor can send the feedback back to the PoC after having reviewed the information carefully. The doctor uses the received data to update the medical history of the patient and can decide if particular actions must be performed from the PoC's side. The classic healthcare system supports the doctor when needed for consulting other experts or completing finer analyses so that the RFID/NFC-PoC is fully integrated with the existing healthcare infrastructure as for the PoC paradigm (see again Fig.~\ref{fig:PoCsHealthcare}). Finally, the doctor's feedback is received by the PoC App and delivered to the patient's side users, too, providing the patient with timely and personalized care. The immediate feedback returned by the RFID/NFC-PoC allows for monitoring of the care even if the doctor's feedback is delayed for any reason.
\subsection{Data Storage, Security, and Representation}
\begin{figure}[t]
    \centering
    \includegraphics[width=\linewidth]{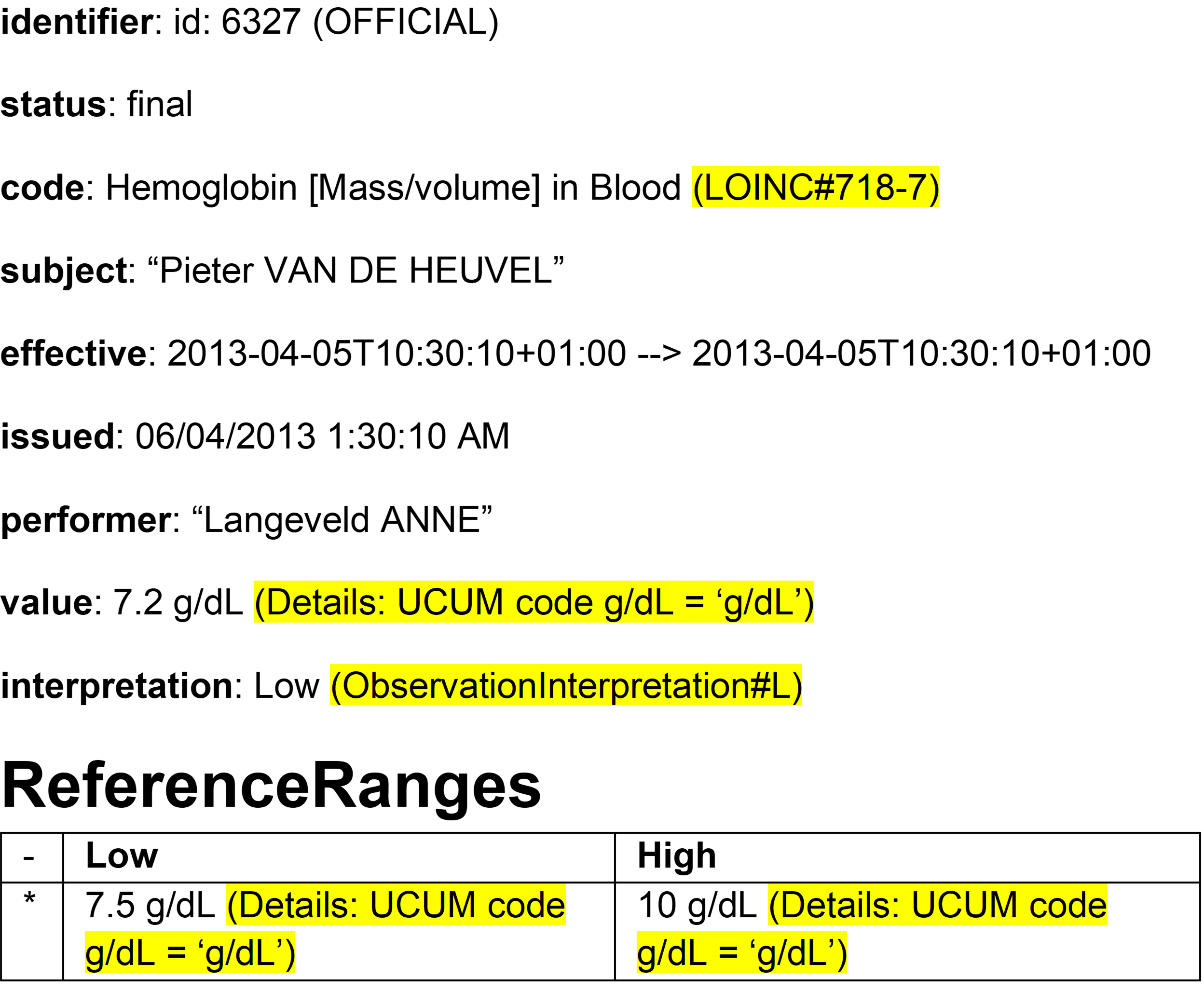}
    \caption{Example of FHIR Observation based on \cite{FHIRObservation}.}
    \label{fig:observation}
\end{figure}
Concerning data security, the tag-reader electromagnetic link has different nature than the reader-doctor link, which exploits the internet. Furthermore, not all data are equally sensitive: asset management information could exploit weaker precautions than vital knowledge like biosignals' measurements.\par
The radiofrequency wireless link is the first link that could be attacked. The security of radiofrequency identification devices has been a research topic since the early 2000s~\cite{MunozAusecha21}, and several reviews have investigated it, including scientometric analyses on research trends~\cite{MunozAusecha21} and existing challenges~\cite{Kumar21}. Several attacks can indeed be performed at this level, for instance, \textit{skimming} through the establishment of a hidden communication link by a malicious reader concealed to the PoC users, or \textit{eavesdropping} on the tag-reader communications, possibly through side-channel attacks~\cite{Raso22Privacy}. The longer the reader-tag distance, the easier performing each attack; thus, the reading range must be as short as possible while the PoC operates. Typical distances of wearable RFID/NFC devices (up to some tens of centimetres) can be hence considered secure if the wearer is vigilant since the attacker should be physically near to the tag and, therefore, easily detectable. The tags must be removed, shielded, or deactivated when they are not expected to be read to avoid skimming. For higher security, tags providing encryption features should be deployed~\cite{Raso22Privacy}.\par
The reader-doctor link is the second one where an attacker can attempt to obtain or manipulate the data, and it poses a cybersecurity vulnerability. Since the amount of data generated by UHF-RFID/NFC sensors is expected to be large, general-purpose Cloud services have to be adopted, but the Cloud service could be an honest-but-curious or malicious actor. Among many possible solutions, ABE can address this issue without requiring complicated key management, as is shown in the next Section. Another point to be addressed in patient-doctor communications is data representation. By exploiting the sophisticated options offered, the FHIR standard plays an extremely important role in this context by standardizing clinical data into files named \textit{Observations}~\cite{saripalle2019using}. Fig.~\ref{fig:observation} shows an example of one FHIR Observation resource. Thanks to FHIR adoption, healthcare information can be seamlessly shared between systems and devices, even between different Nations and languages.
%
%
%
\section{Secure Data Sharing by PoC App}\label{sec:app}
In this Section, a PoC web app for securing the reader-doctor link using ABE is described and implemented.

\subsection{Attribute-based Sharing}

With ABE technology, the patient's side user is able to choose which attributes the person decoding the information must have. For example, he/she can decide that a certain file is readable by all medical and para-medical personnel. In this way, he/she will not be forced to enter the specific credentials of each doctor (like in the case of the classic public key infrastructure solution), which are often not known when the document is shared.

Thus, the following three roles are clearly defined:
\begin{enumerate}
\item the \textit{data owner}, who decides the types of users who can access his data;
\item the \textit{key manager}, typically the hospital or public health services, that provides their employees with credentials in which the attributes characterising them are embedded;
\item a \textit{transport and notification system} that, thanks to the adoption of the ABE encryption technique, operates by ignoring the information being carried or shared.
\end{enumerate}

\subsection{Data Representation}

In order to ease the data exchange and interoperability between UHF RFID and NFC, once data are collected from the tags, the device piloting the RFID/NFC reader converts them into FHIR data. In particular, data collected from the sensors are used to build Observation resources, commonly used to handle, among others, vital signs (e.g., body weight, blood pressure, and temperature), laboratory data (e.g., blood glucose) and device measurements (e.g., EKG data or pulse oximetry data).
\subsection{PoC App Implementation}

\subsubsection{App Overview}

\begin{figure}
    \centering
    \includegraphics[width=\linewidth]{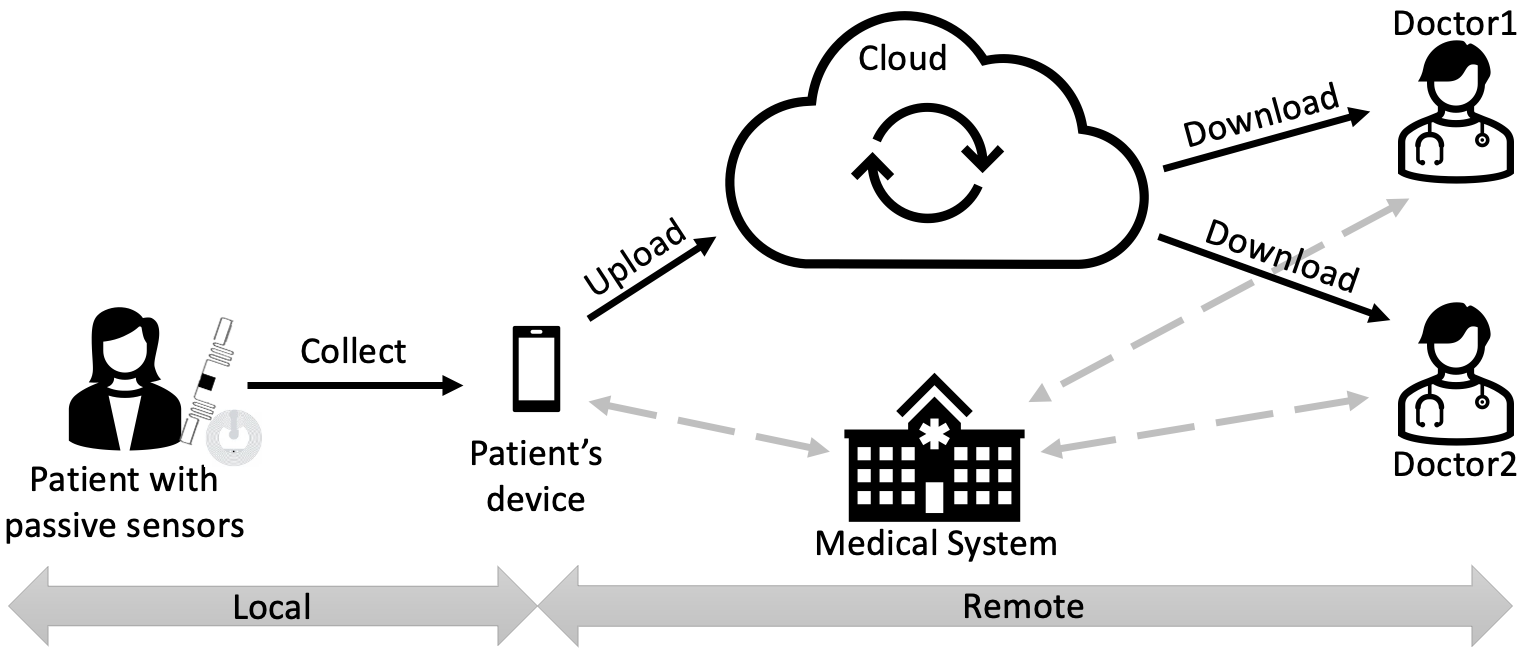}
    \caption{Secure data sharing architecture. Image adapted from \cite{Raso22Privacy}.}
    \label{fig:sec_share_arch}
\end{figure}

We implemented the secure data sharing architecture proposed in~\cite{Raso22Privacy} that adopts the ABE cryptographic scheme to enforce \textit{user-controlled access} (Fig.~\ref{fig:sec_share_arch}). In the remote part of the architecture, which is related to secure Cloud-based data sharing, we consider four actors:
\begin{enumerate}
    \item the {\em Cloud Provider}, one of the existing commercial providers which offer file storage, sharing and synchronisation service (e.g., Dropbox, Google Drive, etc.);
    \item the {\em Patient}, who uses her smartphone or laptop as the reader to collect data from the tags and has to be able to share them with the Medical Personnel;
    \item the {\em Medical Personnel} (or {\em Staff}), whose members have to be able to access data shared by the Patients;
    \item the {\em Medical System}, the entity responsible for the management of the authorisations of the Medical Personnel to access the Patient's data.
\end{enumerate}

Patients and Medical Personnel interact with the Medical System to obtain the cryptographic keys and use them to share protected data using the service offered by the Cloud Provider.

\subsubsection{Technology Details}

\begin{figure}
    \centering
    \includegraphics[width=\linewidth]{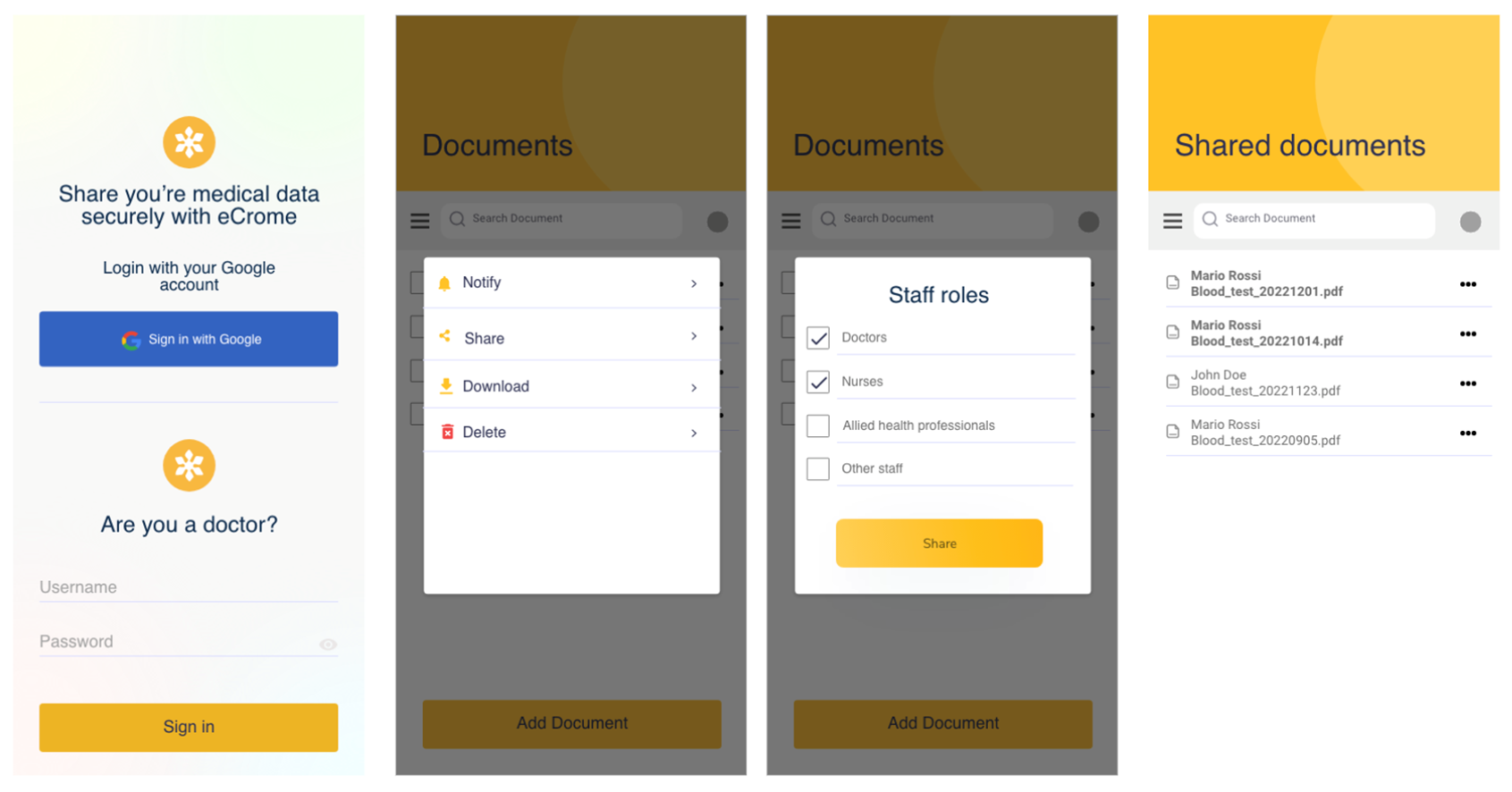}\\
    (a)\qquad\qquad\quad(b)\qquad\qquad\qquad(c)\qquad\qquad\quad(d)
    \caption{Web app screenshots. (a)~Patient/Doctor login. (b)~Sharing menu. (c)~Attribute selection. (d)~Shared document list.}
    \label{fig:web_app}
\end{figure}

We devised the JavaScript Web Application that is named eCrome, which is used by both the doctor and the patient. The main screenshots of this application are shown in Fig.~\ref{fig:web_app}. The application communicates with the Cloud (Google Drive, in this case) to store the patient's documents and share them with the doctors. The protection of the documents that are shared is achieved by means of Rust RABE~\cite{RUSTRABE}, an ABE library written in Rust which was suitably modified and compiled in assembly to be used within browsers. Indeed, to implement the web app, we developed a JavaScript library that wraps RABE to make ABE encryption and decryption possible in a web context. All the complexity is hidden from users, and the cryptographic mechanisms are transparent.

We also implemented two network services: \textit{i})~a key management system that provides the system's standard attributes, the public key ABE and, via login, the private keys of the doctors or health personnel; \textit{ii})~a document sharing notification system based on the no-backend Firebase system~\cite{Firebase} that allows all the web apps to communicate in real-time and thus create document sharing notifications between patients and doctors or between the doctors themselves.
\section{Example of System Implementation}\label{sec:implementation}
An implementation example of the PoC platform is detailed in this Section. The architecture is implemented according to the following steps: \textit{i})~the target application is defined to derive the measurands to be monitored, \textit{ii})~suitable sensors are chosen, \textit{iii})~the radiofrequency identification layer is selected, and \textit{iv})~data security is addressed by utilizing the PoC App presented in the previous Section. Finally, the implemented system is preliminarily tested involving a healthy volunteer.
\subsection{Target PoC Treatment and Measurands Individuation}
The first step necessary to deploy a radiofrequency PoC is defining the quantities of interest based on the medical condition of the patient. Among several possible phenomena, skin temperature and sweat's pH and cortisol levels are helpful for monitoring many illnesses like post-traumatic stress disorder~\cite{Murray21}, dementia~\cite{Pistollato16}, and anorexia~\cite{Winston12}. Without any loss of generality, the implementation example in the remainder of this Section assumes the supervision of the fitness routine of an acute myocardial infarction (AMI) survivor.\par
AMI affects more than $7$ million individuals worldwide, yielding an economic impact of $450$\$ in the United States because of the direct costs solely~\cite{Reed17}. Patients admitted with AMI can also develop subsequent major adverse cardiovascular events (MACE), especially when the cortisol levels in the blood are high~\cite{Jutla13}. Since cortisol increases under stress~\cite{verdejo15,Dahlgren05,pruessner99} in blood and in sweat~\cite{torrente2020investigation}, keeping stress levels low is vital to prevent further MACE events. Physical activity is particularly effective for this aim~\cite{Puterman11, LeBouthillier16}. Nonetheless, physical exercise is not always advisable for survivors of myocardial infarction as the body-fluid balance could be more difficult to maintain than it is for healthy people, possibly increasing the patient's fatigue and stress~\cite{Kavanagh74}. Furthermore, post-traumatic stress disorder following AMI is worryingly common after an invasive intervention~\cite{Cao21}, so the health status can be complicated by hyperventilation~\cite{Foreman95} and the consequent respiratory alkalosis~\cite{Burry01}. Overall, the stress level and physical activity should be closely monitored to ensure the well-being of myocardial infarction survivors.\par
\begin{figure}[t]
    \centering 
    \includegraphics[width=\columnwidth]{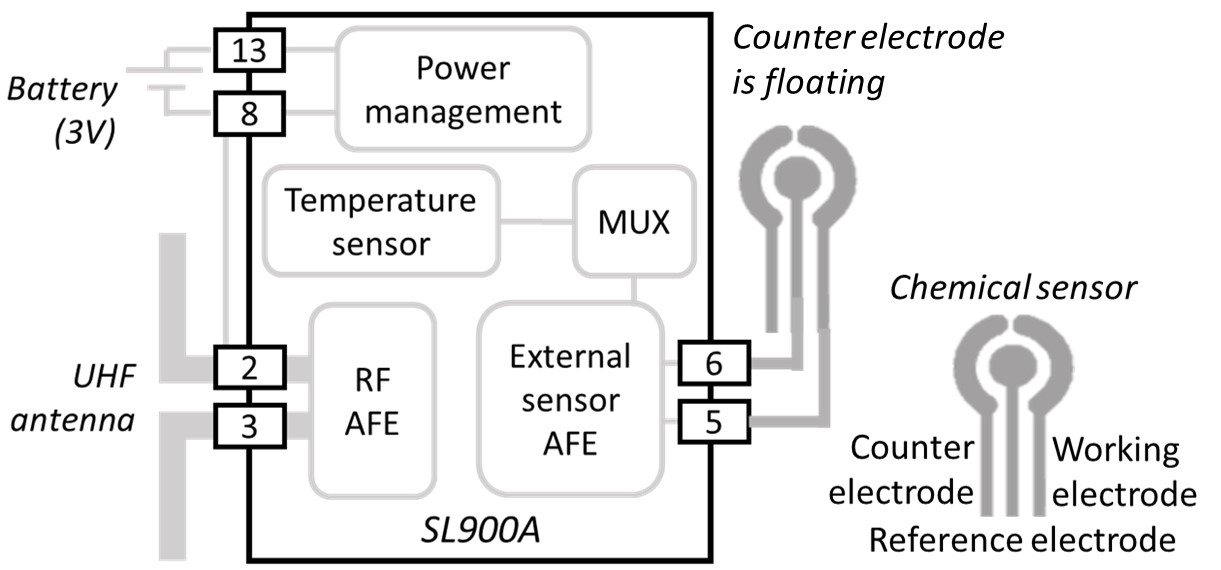}\\
    (a)\\\includegraphics[width=\columnwidth]{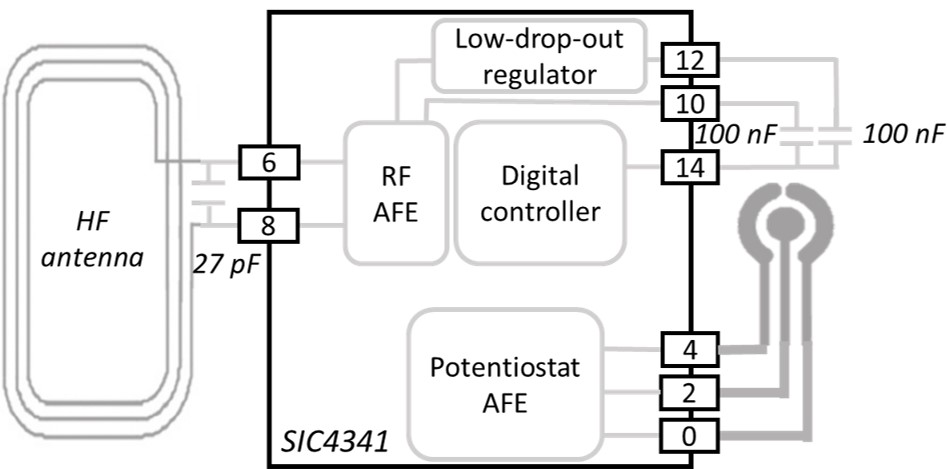}\\(b)
    \caption{Electrical connections of the ICs deployed for the system. Pin numbers and connections are reported, as well as the connections with internal components like an analog front-end~(AFE) and a multiplexer~(MUX). (a)~Schematic of the UHF board with the SL$900$A IC and symbol of the chemical sensor detailing the electrodes. (b)~Schematic of the epidermal NFC sensor hosting the SIC$4341$ IC.\label{Fig:schematicIC}}
\end{figure}
\subsection{Sensors Selection}
%
The domestic PoC platform has to aid an AMI survivor in complying with an exercise routine. Such a platform exploits all the three types of sensing introduced in Section~\ref{sec:introduction}, namely, physical (specifically, temperature), behavioural, and chemical sensing. A temperature sensor monitors the skin temperature of the exercising patient in real-time. During physical activity, indeed, the skin temperature is expected to rise slowly as a consequence of the effort. The physician can monitor the regularity of the fitness routine based on a time log of the read tags and the information on the skin temperature. Two biomarkers in sweat were also chosen to be monitored through chemical sensors: pH, to avoid respiratory alkalosis~\cite{Burry01}, and cortisol, to check if the stress level of the AMI survivor is low enough after the daily routine~\cite{torrente2020investigation,kinnamon2017portable}.\par
The detection of pH in sweat is performed by potentiometric technique, exploiting a modified screen-printed electrode system~(presented in~\cite{Mazzaracchio21}). 
The working electrode was modified with electrodeposited iridium oxide film for having a pH-sensitive layer. The detection of cortisol in sweat, instead, is performed by amperometric detection through a competitive immunosensor~(presented in~\cite{Fiore23}). To complete the cortisol measurement, paper-based microfluidics was used because of its properties of capillary-driven microfluidics and the storing of needed reagents. In this way, the electrode detects cortisol in sweat by simply folding a paper pad close to the electrode and adding a few drops of buffer solution. All the chosen sensors were compared with state-of-the-art benchmarks in previous works, i.e., references \cite{Camera20}~(temperature), \cite{Mazzaracchio21}~(pH), and \cite{Fiore23}~(cortisol).
\subsection{Selected Radiofrequency Identification Layer}
\begin{figure}[t]
    \centering 
    \includegraphics[height=3.5cm]{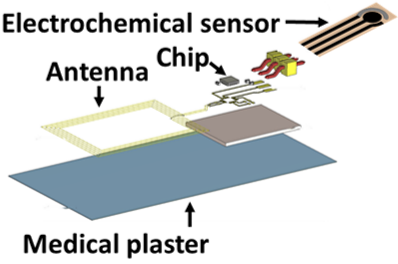} \quad \includegraphics[height=3.5cm]{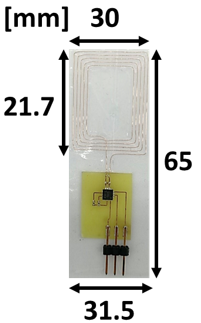}\\
    \qquad (a) \qquad  \qquad \qquad \qquad \qquad \qquad \quad (b)
    \includegraphics[height=3.5cm]{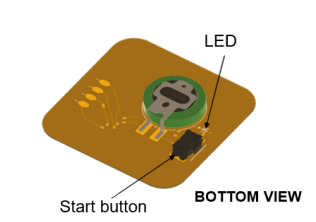}\includegraphics[height=3.5cm]{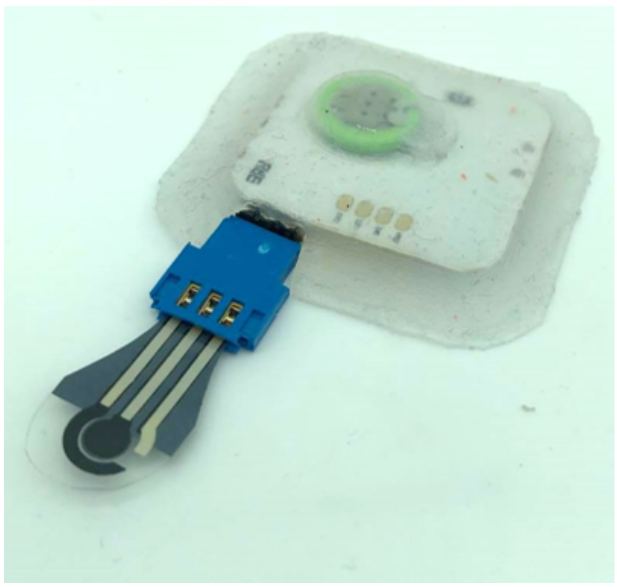}\\
    \qquad (c) \qquad \qquad \qquad \qquad \qquad \qquad (d) \\
    \caption{Hardware utilized for the implementation of the system. (a) Simulated model of the NFC epidermal responder and (b) one realized prototype. UHF RFID epidermal board: top view of (c) the numerical model and (d) one realized prototype. Images adapted from~\cite{Barba22, Nappi21}. \label{Fig:hardwareImplementation}}
\end{figure}
\begin{figure}[p]
    \centering
    \includegraphics[height=4.5cm]{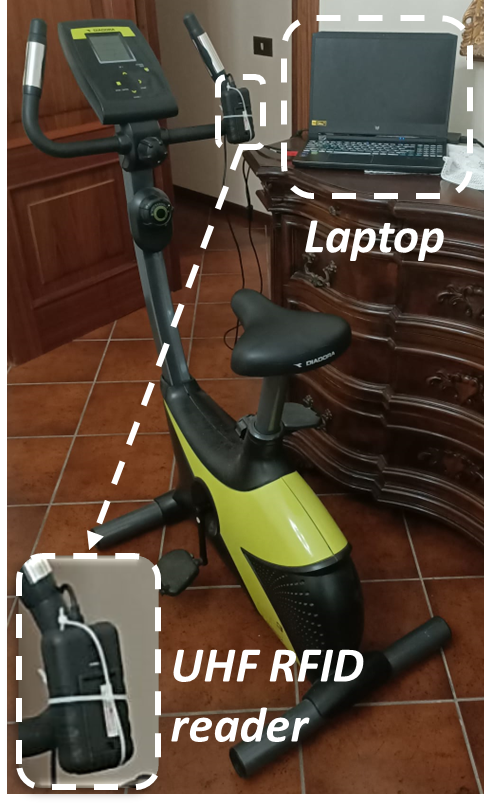}
    \includegraphics[height=4.5cm]{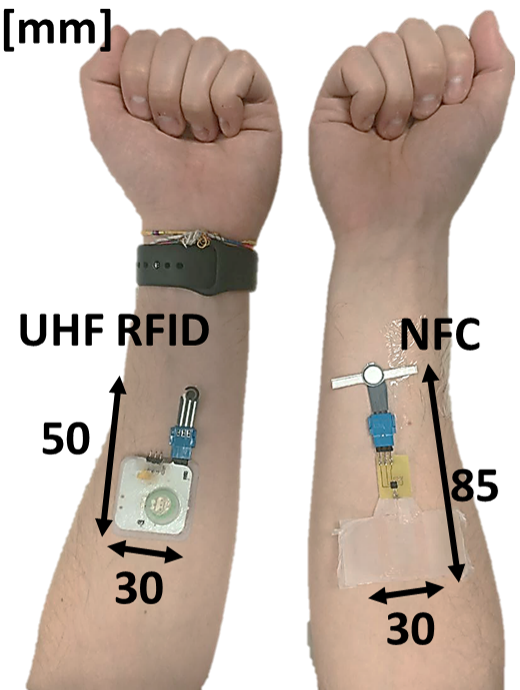}\includegraphics[height=4.5cm]{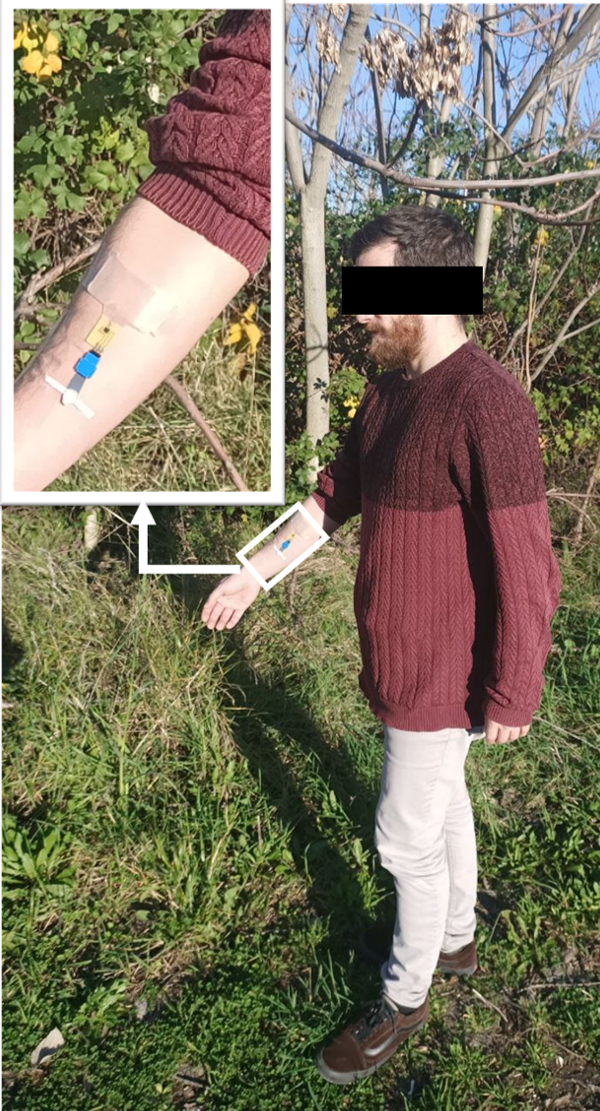}\\
    (a) \qquad \qquad \qquad (b) \qquad \qquad \qquad (c)
    \caption{Set up for the experiments. (a) Stationary bicycle with the USB Plus+ UHF RFID reader. The volunteer who (b) wears both the UHF and NFC sensors, and (c) walks wearing the cortisol sensor (in the inset: zoomed-in view of the cortisol NFC sensor).}
    \label{fig:setUp}
\end{figure}
\begin{figure}[p]
    \centering
    \includegraphics[width=\linewidth]{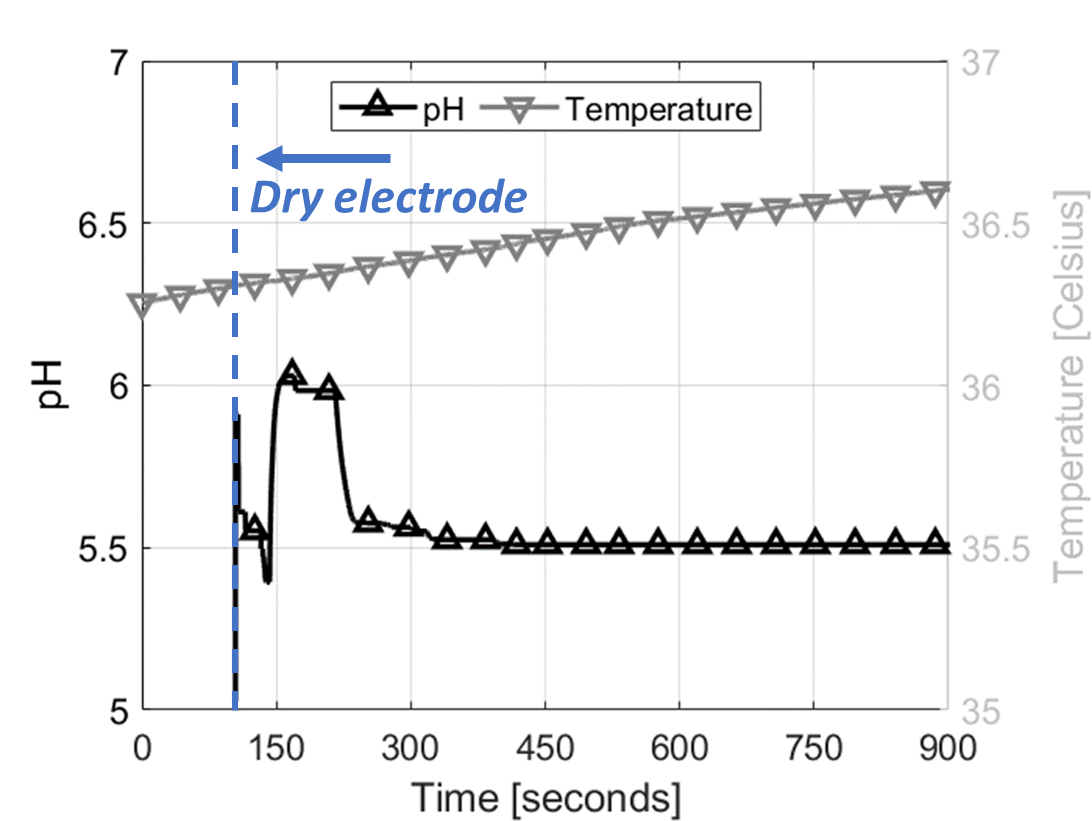}
    \caption{Skin temperature and sweat pH measured by the board during the $15$~minutes of exercise using a stationary bike.}
    \label{fig:measuresPH}
\end{figure}
\begin{figure}[p]
    \centering
    \includegraphics[width=\linewidth]{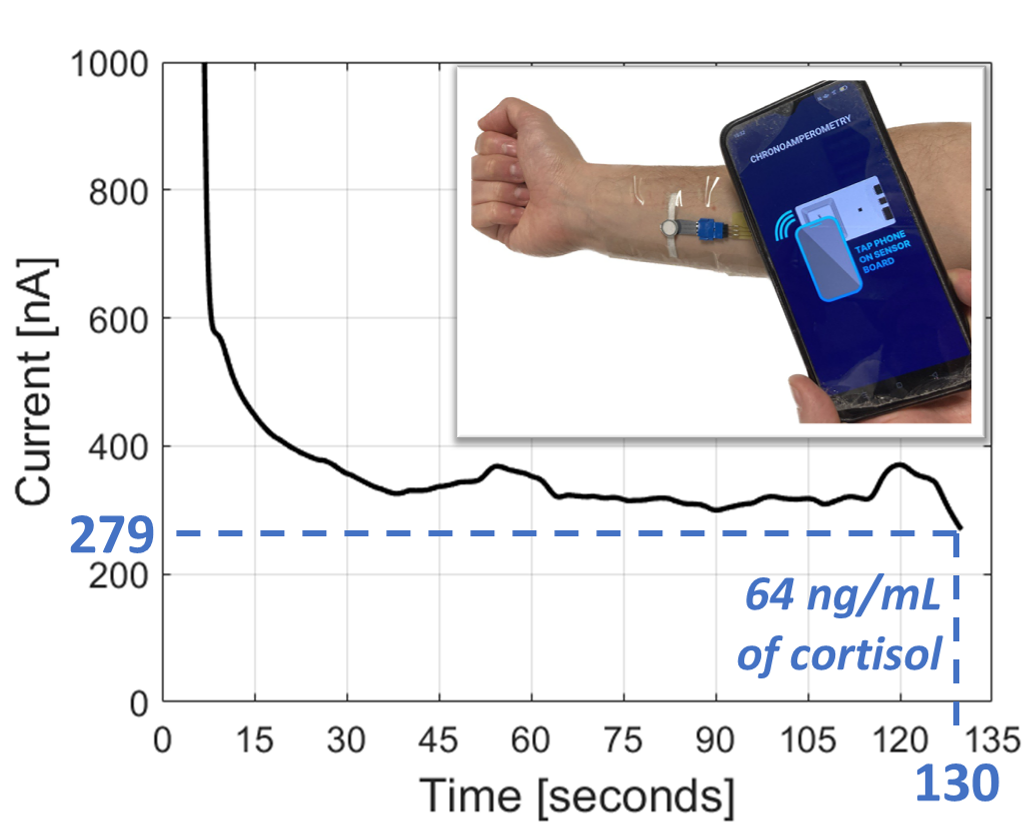}
    \caption{Chronoamperometric cortisol sensing after the $15$-minute-long walk. Time of measurement, current measured and cortisol level are reported. In the inset: chronoamperometry performed through the NFC board and the Chemister app.}
    \label{fig:measuresCortisol}
\end{figure}
\begin{figure*}   \centering
    \includegraphics[width=5.9cm]{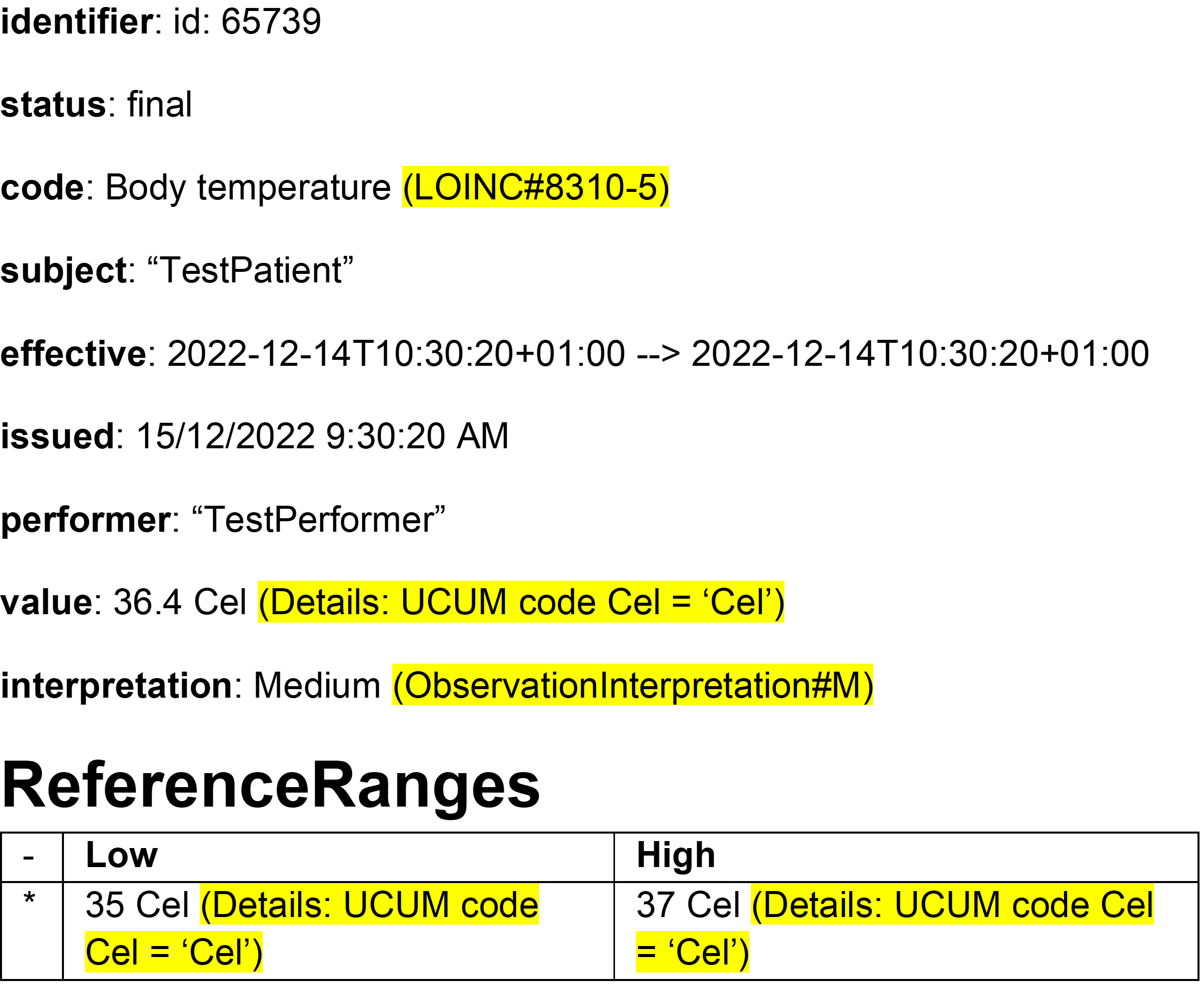}
    \includegraphics[width=5.9cm]{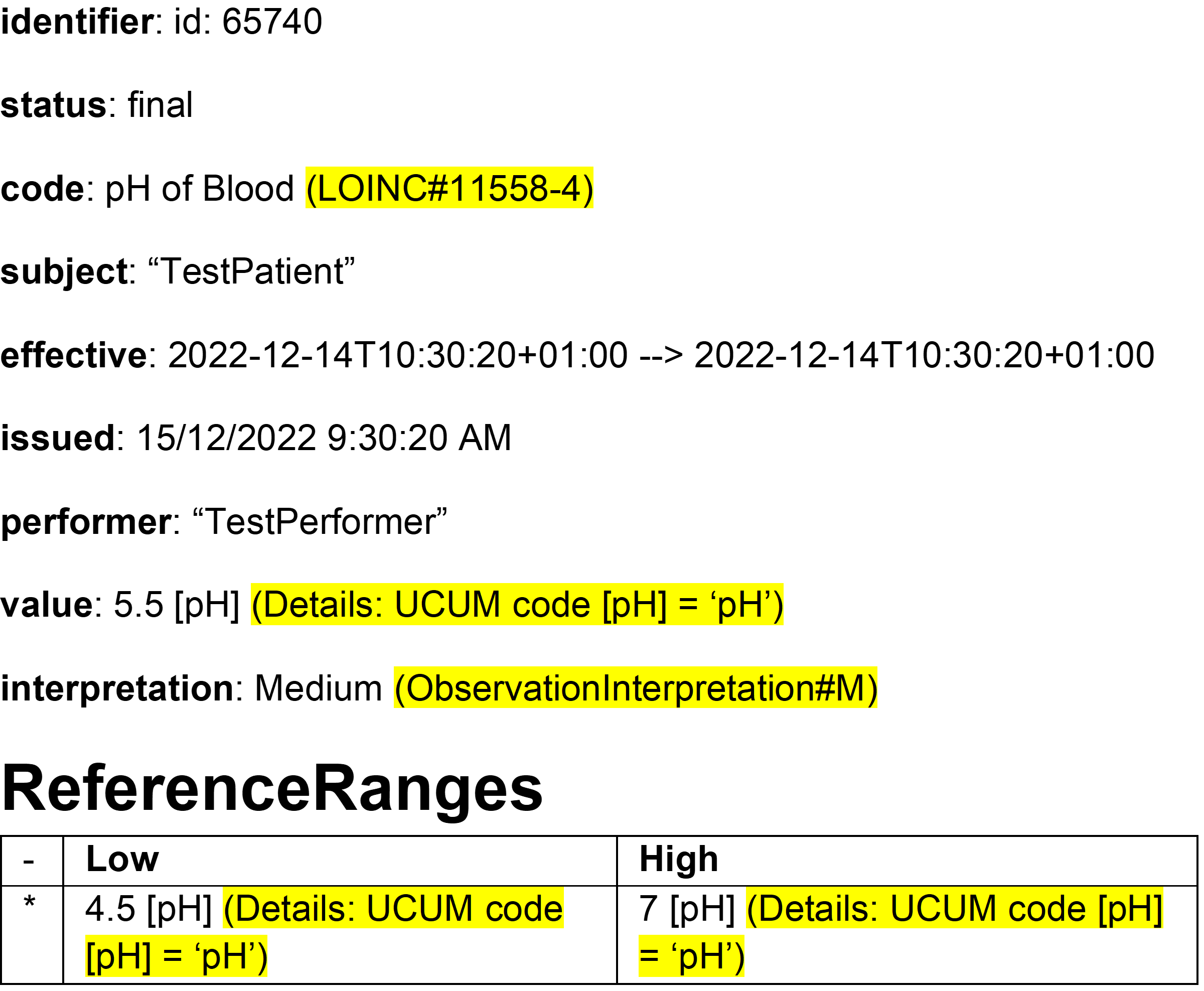}
    \includegraphics[width=5.9cm]{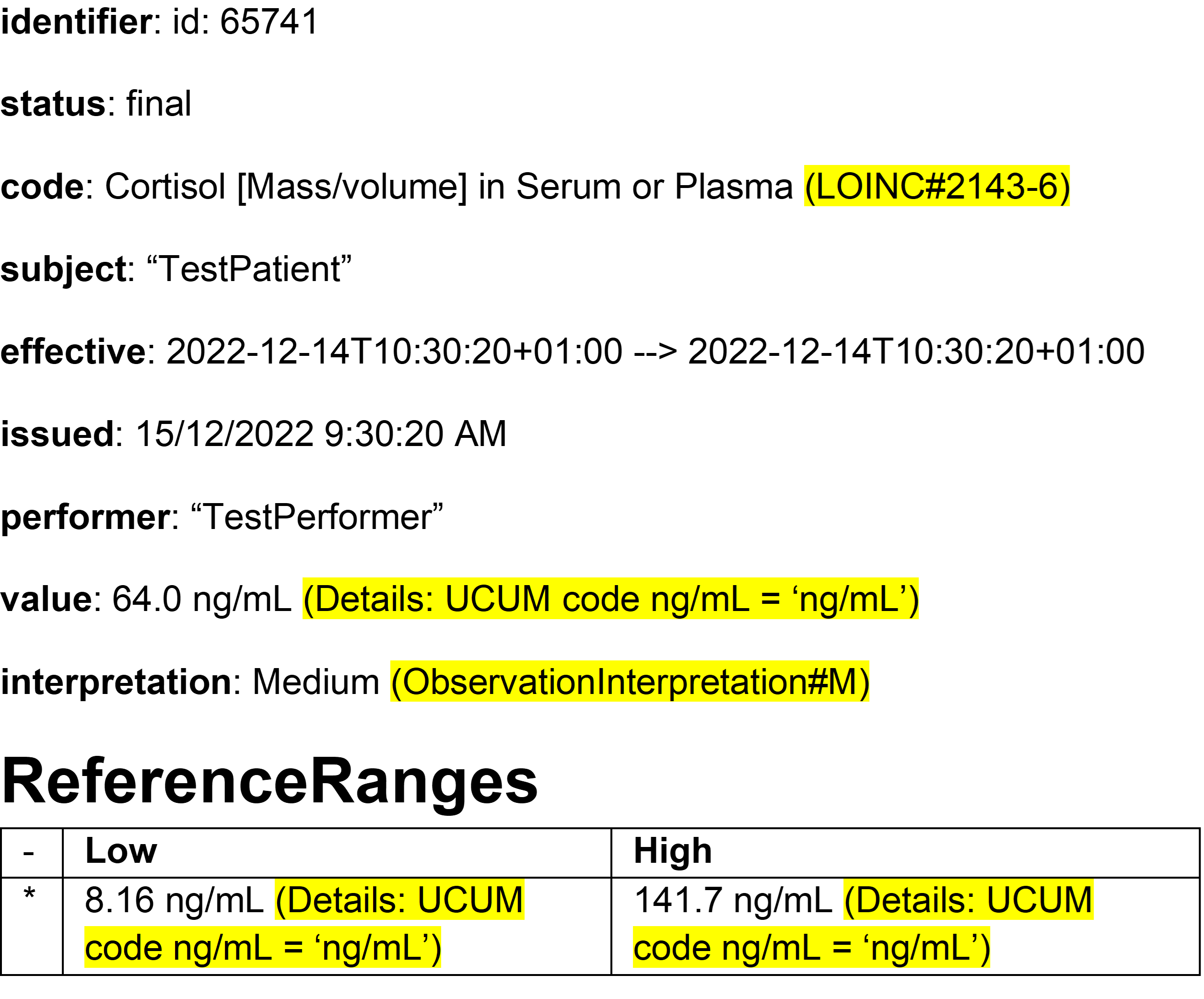}\\
    (a) \qquad\qquad\qquad\qquad\qquad\qquad\qquad\qquad (b) \qquad\qquad\qquad\qquad\qquad\qquad\qquad\qquad (c)
    \caption{FHIR observation of the performed measurements of (a) temperature, (b) pH, and (c) cortisol.}
    \label{fig:observationTests}
\end{figure*}
The radiofrequency components were selected based on the pros and cons of the two technologies recalled in Section~\ref{sec:introduction}. Since cortisol level varies slowly along the day~\cite{Smyth97}, the level of cortisol in the sweat should be measured using an NFC epidermal device in order to \textit{i})~avoid the use of any battery to perform the power-hungry chronoamperometry, \textit{ii})~ensure the maximum possible security of the data, and \textit{iii})~make it possible to interrogate the tag in any time and place through the smartphone. Instead, physical activity needs longer reading distances to give the wearer some freedom of movement in the proximity of the reader during real-time monitoring and, accordingly, a UHF RFID board is the optimal choice. Hence, the electrochemical sensors will be connected to the proper electromagnetic interfaces as needed, and a hybrid UHF-NFC system will be implemented. \par
The NFC responder from~\cite{Barba22} and the epidermal UHF board from~\cite{Nappi21} can be used as body-worn epidermal tags. Fig.~\ref{Fig:schematicIC} reports the schematics of the two tags, whereas the numerical models and prototypes are depicted in Fig.~\ref{Fig:hardwareImplementation}. The tags can be worn for up to some hours as required for monitoring physical activity. The NFC coil is manufactured through $40$-$\mu$m-wires manually posed on a breathable and transparent plaster (Tegaderm by 3M\textsuperscript{TM}). The spiral antenna is soldered to the SIC~$4341$ (from Silicon Craft Technology) IC that can be read by smartphones through the ISO/IEC~$14443-3$A protocol while performing biosensing. An FR-$4$ pad hosts the IC SIC~$4341$ and the plug\&play connector for the sake of robustness. The chosen NFC reader embedded in a smartphone (Oppo Reno Z; operating system: ColorOs V11.1) did not show any interoperability issue with the NFC sensor, reading it smoothly.\par
The RFID tag has the SL$900$A (by AMS OSRAM) IC, which is used in battery-assisted-passive (i.e., the battery is utilized for power but not for communications) mode to lower the chip sensitivity down to~$-15$~dBm. The antenna is an open-loop (maximum gain: $-15$~dBi). The board can be read by the portable UHF RFID ``USB Plus+" (by ThingMagic) reader having an embedded antenna and maximum equivalent isotropic radiated power of $24$~dBiL resulting in a maximum reading distance of about $15$~cm with the selected tag~\cite{Nappi21}. This distance allows the board's wearer to ride the stationary bicycle comfortably while using the tag. Then, the USB Plus+ reader can be connected to a piloting laptop and fixed to the stationary bicycle~[Fig.~\ref{fig:setUp}(a)]. In this way, the tag-reader link is the shorter and more secure possible, and the European regulation on the specific absorption rate is respected as the arrangement is similar to the one described in~\cite{Miozzi19Constrained}.\par
%
%
\subsection{Use of the PoC Platform}
In this example, the doctor is assumed to prescribe to the AMI survivor $15$~minutes of exercise on a stationary bike as the daily fitness routine while monitoring the skin's temperature and the pH of the sweat to avoid respiratory alkalosis. Afterwards, the survivor should walk for additional $15$ minutes and, at the end of the routine, check the cortisol level in sweat to inquire if it is in the desired range. The doctor can hence check compliance with the fitness routine based on the timestamps, record the medical data, and monitor the overall psycho-physical well-being of the patient through the PoC App.\par
The radiofrequency boards were attached to the right and left ventral mid forearms of a healthy volunteer simulating the AMI survivor~[Fig.~\ref{fig:setUp}(b)]. Ventral mid forearms were chosen as the application points of the sensors since they are among the optimal positions for the targeted sweat sensing~\cite{Baker18}. Afterwards, the volunteer walked in a park at his usual walking speed~[Fig.~\ref{fig:setUp}(c)]. All the retrieved raw data collected by the laptop and the smartphone were given as input to the PoC App, which post-processed them by discarding incorrect readings and performing a moving average to smooth fluctuations due to displacements of the sensors. Accordingly, pH values of the sweat comprised between $4.5$ and $7.0$ were considered correct based on \cite{Baker18}, and the window for averaging was of $5$ seconds.\par
The recorded tracks of pH and temperature are drawn in Fig.~\ref{fig:measuresPH}. During the exercise, the skin temperature increased monotonically by about $0.5$~Celsius, as expected, suggesting a continuous and physiological physical effort. pH monitoring started after about $100$~s of cycling when the volunteer started sweating, and the sensing electrode got wet. The pH value stabilized at $5.5$ after $300$~s, confirming that the subject is healthy and no adverse event happened during the exercise. After the walk, the cortisol level was checked through chronoamperometry. The current value, which stabilizes after $130$~s, was obtained by the Chemister app (by Silicon Craft Technology) and then given to the PoC App. After performing the conversion, the measured cortisol in sweat was $64$~ng/mL (Fig.~\ref{fig:measuresCortisol}), fully comparable with physiological cortisol levels of healthy subjects reported by the literature, namely, $8.16-141.7$~ng/mL post-exercise~\cite{Russell14}. Even in this case, the NFC board confirmed that the subject was healthy and not excessively stressed, as expected. Finally, data were converted into FHIR observations (Fig.~\ref{fig:observationTests}) and transmitted through the PoC App to another device emulating the doctor's smartphone.
\section{Conclusion}\label{sec:conclusion}
In this paper, we proposed a point-of-care platform utilizing UHF RFID and NFC devices for the secure collection and transmission of medical data. The high-level architecture of the platform, the hardware selection, and data security and representation were analyzed. Data from the two technologies are made homogeneous according to the FHIR healthcare standard and are securely shared by a web app utilizing Cloud storage and ABE. An implementation example regarding sensing cortisol, sweat's pH, and skin temperature through the hybrid platform was deployed and tested.\par
The architecture and the implementation presented above confirm that thanks to the combination of UHF RFID and NFC, the latest advancements in radiofrequency sensors and healthcare information management can effectively be integrated into fully functioning PoCs that can address the severe challenges faced by healthcare systems worldwide. The investigation proves the concept and the feasibility of this kind of points-of-care that yet need to be tested in real case studies to quantify the expected benefits exactly. However, using the two radiofrequency identification technologies together is still difficult since it requires a complex system architecture. The development of new sensing dual-tags~\cite{Rizzi19} and the latest generation of platforms embedding both NFC and UHF readers~\cite{C66} can significantly simplify hardware deployment. Furthermore, the tag-reader links cannot be considered secure if long reading distances are exploited. Current signs of progress on reconfigurable wearable metasurfaces could help ensure security and privacy~\cite{Li21}, for instance, by changing the radiation pattern of the body-worn antenna to hinder eavesdropping.
\section*{Acknowledgments}
The authors thank Dr Carolina Miozzi and Ms Adina Bianca Barba (from RADIO6ENSE srl), and Ms Alessia Riente (from the Pervasive Electromagnetics Lab of the Tor Vergata University of Rome) for their valuable help in completing the implementation example. \par
\bibliographystyle{IEEEtran}
\bibliography{main_rfidj.bib}
\end{document}